\documentclass[proof]{WileyASNA-v2}

\articletype{Original Article}%

\received{X X 2018}
\revised{X X 2018}
\accepted{X X 2018}

\begin{document}

\title{A catalogue of spectroscopic binary candidate stars derived from a comparison of Gaia DR2 with other radial velocity catalogues}

\author[1,2]{D. Jack*}
\authormark{D. Jack}

\address[1]{\orgdiv{Departamento de Astronom\'\i{}a}, \orgname{Universidad de Guanajuato}, \orgaddress{\state{Guanajuato}, \country{Mexico}}}
\address[2]{\orgdiv{Hamburger Sternwarte}, \orgname{Universit\"at Hamburg}, \orgaddress{\state{Hamburg}, \country{Germany}}}

\corres{*D. Jack, . \email{dennis@astro.ugto.mx}}

\presentaddress{Departamento de Astronom\'\i{}a, Universidad de Guanajuato, A.P.~144, 36000 Guanajuato, GTO, Mexico}

\abstract{Using the recently published Gaia second data release which includes measurements of the mean radial velocity of about 7.2 million stars, we performed a systematic comparison
with other existing radial velocity catalogues in order to search for variations in the radial velocity measurements, with the goal that
detected differences may indicate that these stars are possibly spectroscopic binaries stars with only one visible component (SB1).
We present a spectroscopic binary candidate stars catalogue containing 35,246 stars, compiled to encourage follow-up observations
obtaining spectra at different epochs of these stars orbits in order to verify their binarity and to study these systems using radial velocity curves.
Comparing the Gaia DR2 database with the K-M dwarf catalogue we found 16 stars that show radial velocity variations.
In a comparison with the Pulkovo radial velocity catalogue of Hipparcos stars we identified a total of 539 SB1 candidate stars. 
In the largest radial velocity catalogue available, the RAVE catalogue, we found a total of 34,691 stars that show radial velocity variations when compared to the Gaia DR2 data.}

\keywords{binaries: spectroscopic, catalogs, techniques: radial velocities}

\fundingInfo{CIIC 2018 DAIP, UG}

\maketitle

\section{Introduction}\label{sec1}

Binary or multiple stellar systems are very common in our Galaxy and our Universe. 
Despite their ubiquity, the detection of such systems is difficult.
There exist several methods for the identification of binary or multiple stellar systems.
Actually, these are the same methods as are used for the detection of exoplanets.
In some binary systems, both stars can be observed directly by imaging as so called visual binaries.
In addition, using astrometric measurements these systems can be studied in detail 
in order to determine
the orbital properties. Astrometry of only
one visible star is another method to detect binary systems.
A large number of visual binaries have been studied, and they have been published, e.g. in The Washington Double Star Catalog \citep{mason01,wdsc}.
Observations monitoring the photometry of stars leads to the discovery of transits and thus,
the detection of eclipsing binary or multiple systems, which need to have an orbital inclination close to 90$^\circ$.
This has been achieved with the Kepler mission resulting in a very large catalogue of eclipsing binary stars \citep{kepler}.

Another successful detection method is the measurement of changes in the radial velocity of a star caused by the movement through its orbit.
However, this method requires spectra of the object at different epochs to reveal radial
velocity variations. In some cases spectral lines of both stars can be observed in the spectra (SB2). However, usually only one
component is visible and a single-lined spectroscopic binary (SB1) is observed.
The analysis of the radial velocity curve reveals information about of the orbital parameters and may provide
implications about the invisible companion star. Also there exist a few catalogues of spectroscopic binary systems and
their orbital parameters like the 9th Catalogue of Spectroscopic Binary Orbits \citep{sb9} or the CHARA Catalog of Spectroscopic Binary Stars \citep{chara}. 
Detecting a spectroscopic binary can either be pure luck or requires a survey
at different epochs like the Gaia-ESO Survey \citep{gilmore12}, 
APOGEE \citep{majewski16,majewski17} or LAMOST \citep{cui12}.
The rate of positive detections is for several reasons quite low ($\approx 10\%$)
\footnote{This is just a rough estimate for a random lucky search. The detection rate in this work is
significantly lower due to selection effects.}.
One selection effect is that binary systems with a low inclination are very hard to detect.
In addition, long period binaries require a longer observational campaign and some extreme eccentric orbits may show radial velocity
variations only during a short part of the curve. The low detection rate also
comes from instrumental limitations, since the precise determination of the radial velocity
requires observations with high spectral resolution.

The Gaia mission \citep{gaiamission} was launched on December 19, 2013. It started observations
in July 2014, and the first data release was published on September 14, 2016
with measurements of the position of more than one billion stars, and in
addition measurements of their proper motions and parallaxes.
Recently, Gaia published its second data release \citep{gaiadr2}, which now
includes measurements of the radial velocity of about 7.2 million stars \citep{gaiarv}.
These published values are only mean values of several measurements of the radial velocity. Since Gaia DR2 is a
very complete database, comparing the Gaia DR2 data with measurements of radial velocities at different epochs 
published in other catalogues
provides the opportunity to find and identify possible SB1 stars.
This will allow a search for spectroscopic binaries and the measurement of their radial velocity
curves for orbit determination.

We compare the Gaia DR2 data with three other radial velocity catalogues.
We describe our general method in some detail in Section~2. Section~3 contains the
results of our comparison and a large catalogue of spectroscopic binary candidate stars. 
A summary and our conclusions are presented in Section 4.

\section{Method}\label{sec2}

There exist several publications and catalogues with measurements of the radial velocity
of a large number of stars. In this work, we  compare the recent publication of the radial velocities
of the Gaia data release 2 with the available data of three other catalogues. 
In this section, we briefly describe 
the general method that we applied in our search for spectroscopic binaries.

The recently published Gaia DR2 for the first time contains measurements
of the radial velocity of some of the sources \citep{gaiarv} obtained with
the Gaia Radial Velocity Spectrometer \citep{gaiarvspectrometer}.
While initially measurements for more than 9.8 million sources had been obtained, after 
a selection process the final data release contains the radial velocities of 
about 7.2 million objects. During the selection process clear spectroscopic binary
stars had been eliminated, which will also lower the detection rate of our search.
This was done for all stars
that show a difference between the two components larger than 20~km\,s$^{-1}$ \citep{sartoretti18}.
However, there will still be a very large amount of undiscovered
spectroscopic binaries among the published data, which usually show a somewhat larger
error for the radial velocity than do single stars.

For our comparison, we used three catalogues: the catalogue for K-M dwarf stars of \citet{kmdwarfs}, 
the Pulkovo catalogue \citep{pulkovo} from Hipparcos stars and the
fifth data release \citep{rave} of the RAVE survey.

The first main obstacle is to correctly identify the stars of each of the catalogues
with the respective source in the Gaia DR2, since it 
does not come with a source identification in other catalogues (like HD etc.)
but only its proper source ids.
However, there already exist two cross match lists of the Hipparcos2 and RAVE DR5 stars
giving their most likely neighbour in the Gaia DR2 database, which we make use of in this work.
For the stars of the K-M dwarf catalogue we searched directly in the Gaia DR2 database
for sources with radial velocity measurements
within a few arcseconds around the position
of the stars given in these catalogues. The size of this search-window depends
on the precision of the position in the catalogue since Gaia has,
for obvious reasons, very precise position measurements.
For the K-M dwarf catalogue, we searched within a window of 10 arcseconds of the same position, since
the coordinates given in this catalogue are not very precise.
Because this catalogue is not very large, double or multiple identifications
are revised by hand to assure the final correct identification of the sources.

The identification process results in a catalogue
of a large number of stars with measured values for the radial velocity in both the
Gaia and the other catalogue respectively. Not all the sources of the other catalogues
also have measurements in Gaia DR2.
The next step of our method is to check
for variability in the radial velocity measurements between the two catalogues.
The selected radial velocity variation criterion depends on the error of the radial velocity in each of
the catalogues. We take a conservative value of a variation of more than 
three times the respective error of the measured values in the catalogues.
As a last step, we eliminate all photometric variable stars.
Note that this general procedure may have to be adapted slightly for each of the three catalogues.
Thus, we present the specific details in the next section where we also present the
results of the search, statistics, and details of the spectroscopic binary candidate stars.

\section{Spectroscopic binary candidate stars}

We applied the above described method to three different catalogues
of radial velocity measurements in order to identify
the stars that show variations with respect to the radial velocities reported in the Gaia DR2 catalogue.
In this section, we explain in detail the results for each of the three catalogues
and present the list of spectroscopic binary candidate stars.

\subsection{Radial velocities of K-M dwarf stars}

The first catalogue that was compared with the Gaia DR2 was
the radial velocity catalogue of K-M dwarfs \citep{kmdwarfs}. This 
catalogue contains radial velocity observations of 959 stars that
are classified as K or M dwarf stars. Only stars
not suspected of having any radial velocity variations were selected,
which will lower the expected detection rate of our search significantly.
However, there is always the possibility that some spectroscopic binaries
are still hidden in this catalogue. We selected this recently published catalogue 
since it has a reasonable number of stars so that the small number of 
obtained spectroscopic binary star candidates can also be checked by hand.

We first performed an automatic identification procedure of the stars of the catalogue with the Gaia DR2 sources
using a search-window of 10 arcseconds. Double or multiple identifications were checked by hand,
and photometric variable stars were eliminated.
We found that a total of 650 stars have been clearly observed in both catalogues.
The radial velocity measurements of the K-M dwarf catalogue have relatively small errors
of about $\approx$ 0.5~km\,s$^{-1}$.
Therefore, we checked for radial velocity differences of more than 4~km\,s$^{-1}$ between the two catalogues
to take into account the errors in both the K-M dwarf and the Gaia DR2 catalogues and
a probable offset between the catalogues.
After applying this criterion, we obtained a list of 16 stars that show radial velocity variations.
This small number clearly indicates that the K-M dwarf catalogue has mostly stars that do not
show radial velocity variations. The detection rate of suspected SB1 stars in 
this catalogue is just 2.46~\%.

\begin{center}
\begin{table*}[t]%
\centering
\caption{List of stars of the K-M dwarf catalogue that show a variation
of more than 4~km\,s$^{-1}$ in the radial velocity measurement when compared to
the values of the Gaia DR2 database.}
\tabcolsep=0pt%
\begin{tabular*}{450pt}{@{\extracolsep\fill}lcccccccc@{\extracolsep\fill}}
\toprule
\textbf{Star} & $m_{\mathrm{V}}$& \textbf{$\bar{v}_\mathrm{K-M}$~(km\,s$^{-1}$)} & $N$ & \textbf{$v_\mathrm{K-M}$~(km\,s$^{-1}$)}  & \textbf{Epoch$_\mathrm{K-M}$ (MJD)} & \textbf{$v_\mathrm{Gaia}$~(km\,s$^{-1}$)}\\
\midrule
GJ 9038 B &  9.85 &   $-0.7 \pm 0.3$ & 6 & $-2.2 \pm 1.1$ & 2455476.536 & $-4.75 \pm 0.84$\\
                   &           &   &    & $-0.5 \pm 0.6$ & 2455478.474	 & \\
                   &            &  &    & $-0.7 \pm 0.6$ & 2455479.449	 & \\
                   &            &  &    & $-1.0 \pm 0.7$ & 2455482.464	 & \\
                   &            &  &    & $ 0.4 \pm 0.9$ & 2455484.461	 & \\
                   &            &  &    & $-0.9 \pm 0.9$ & 2456681.221	 & \\
HIP 14165 & 11.792 &  $21.8 \pm 0.7$ & 1 & $21.8 \pm 0.7$ & 2452930.914 &  $33.82\pm 7.03$\\
HIP 14593 & 10.06 &   $28.2 \pm 1.2$ &  1 & $28.2 \pm 1.2$ & 2452928.894 &  $15.93 \pm 0.48$\\
HIP 19948 & 10.901 &$48.0 \pm 0.8$  & 1 & $48.0 \pm 0.8$ & 2452330.632 &  $42.26 \pm 1.12$\\
HIP 34361 & 11.081 & $-16.8 \pm 0.7$ &  1& $-16.8 \pm 0.7$ & 2452328.721	 &  $26.98 \pm 0.35$\\
HIP 40724 & 9.868 &   $88.1 \pm 1.0$  & 2 &  $89.2 \pm 1.0$ & 2455881.616	 &  $81.9 \pm 1.87$\\
                   &            &  &   &  $87.1 \pm 1.0$ & 2456679.455	&  \\
TYC 1379-91-1 & 11.63 &$29.4 \pm 0.4$  & 4 & $30.0 \pm 0.7$ & 2455271.365 &  $-24.26 \pm 0.69$\\
                          &        &   &     & $29.3 \pm 0.7$ & 2455278.320 &  \\
                          &         &  &     & $28.6 \pm 0.9$ & 2455302.275 &  \\
                          &         &  &     & $29.5 \pm 0.8$ & 2455969.495 &  \\
HIP 44109  & 10.360& $26.9 \pm 0.5$ & 1 & $26.9 \pm 0.5$ & 2455618.402 &  $22.6 \pm 6.86$\\
HIP 46662  & 10.09 & $-51.3 \pm 0.3$  & 3 &  $-50.8 \pm 0.5$ & 2451622.728 &  $-45.22 \pm 1.24$\\
                    &          &  &    &  $-51.3 \pm 0.5$ &    2455266.372        & \\
                    &          &  &    &  $-51.7 \pm 0.5$ &    2455655.361        & \\
HIP 49969  & 10.629 &  $25.9 \pm 0.6$ & 2 & $26.1 \pm 0.8$ & 2455621.430 &  $31.59 \pm 1.28$\\
                    &             & &    & $25.7 \pm 0.8$ & 2455655.323 &  \\
HIP 58099  & 10.5 & $-19.0 \pm 0.5$ & 2 &$-18.7 \pm 0.7$ & 2455610.611 &  $-14.88 \pm 0.8$\\
                    &         &  &   &$-19.2 \pm 0.6$ & 2455655.508 &  \\
HIP 59000  & 10.016 &  $-8.2 \pm 0.4$& 1 & $-8.2 \pm 0.4$ &  2452328.912 &  $-29.16 \pm 13.6$\\
TYC 4558-1019-1 &  10.97 &  $-102.0\pm 0.3 $  & 3 & $-101.6\pm 0.4 $ & 2452349.524 &  $-19.22 \pm 1.54$\\
                                &          &  &  & $-102.3\pm 0.5 $ & 2455352.364 &  \\
                                &          &  &  & $-102.5\pm 0.5 $ & 2455622.566 &  \\
GJ 4327  & 12.07&  $13.4 \pm 1.2$ & 2 & $14.0 \pm 0.9$  & 2452927.796& $-6.68 \pm 2.23$\\
                &          &  &    & $11.0 \pm 1.8$  & 2455445.571 &  \\
HIP 115004 & 9.75 & $26.6 \pm 0.4$ & 3 &  $27.1 \pm  0.4$ & 2455497.350 &  $19.13\pm 0.87$\\
                    &          & &    &  $26.7 \pm 0.4$ & 2455777.538 &  \\
                    &          & &    &  $25.5 \pm 0.5$ & 2455836.444 &  \\
HIP 116936 &  10.293&$17.7\pm 0.3$  & 3&$17.0\pm 0.8$ & 2455473.467 &  $28.43\pm 2.37$\\
                    &              &  &   &$18.1\pm 0.6$ & 2455853.374 &  \\
                    &              &  &   &$17.6\pm 0.6$ & 2455881.275 &  \\
\bottomrule
\end{tabular*}
\label{tabkm}
\end{table*}
\end{center}

In Table \ref{tabkm}, we present the list of 16 stars that show radial velocity variations
of more than 4~km\,s$^{-1}$. The table is sorted by right ascension coordinate as in
the K-M dwarf catalogue. The second column gives the magnitude in the $V$-band reported in
the SIMBAD database. The third column gives the mean radial velocity of the K-M dwarf
catalogue ($\bar{v}_\mathrm{K-M}$ in km\,s$^{-1}$) while the fourth column gives  the number of individual observations ($N$) used for
 the radial velocity determination.
The fifth column contains all individual measurements of the radial velocity ($v_\mathrm{K-M}$ in km\,s$^{-1}$),
and the respective epochs are given in column six.
The last column contains the measurements and errors of the radial velocity in the Gaia DR2 ($v_\mathrm{Gaia}$ in km\,s$^{-1}$).
Since these are only a few spectroscopic binary candidate stars, we will briefly discuss each one of them
in detail in \ref{sec:km}. 
We also looked into the literature for additional publications of the radial velocity of
these stars.

We found that one star, namely HIP~19948, has already been
classified as a double or multiple star system in the SIMBAD database \citep{simbad}. 
This star is actually the visual binary WDS J04167-1233AB listed in the Washington Double Star
Catalog \citep{mason01,wdsc} with a separation of 2.5 arcseconds between the two components.
With a variation in the radial velocity of about 5.7~km\,s$^{-1}$ and 
an error of the Gaia DR2 measurement larger than 1~km\,s$^{-1}$ the star could also be a spectroscopic binary.

%
\subsection{Pulkovo compilation of radial velocities for Hipparcos stars catalogue}

The second catalogue that we compared with the Gaia DR2 radial velocity measurements
is the Pulkovo compilation of radial velocities for Hipparcos stars \citep{pulkovo}
containing published radial velocity measurements of 35,495 Hipparcos stars.
The stars of this catalogue have different spectral type and luminosity
class but are located within a sphere of 500~pc centered in the Sun having known parallaxes obtained
by Hipparcos \citep{hipparcos1,hipparcos2}.
This catalogue is a compilation of radial velocities reported in 203 publications. The
median accuracy is roughly 0.7~km\,s$^{-1}$, but some measurements have errors of up to 5~km\,s$^{-1}$.

We used the Hipparcos2 - Gaia DR2 cross match list to check for Gaia DR2 sources that have radial velocity measurement 
in both catalogues and obtained a final list of 16,601 stars.
In the next step, we searched for variations in the radial velocity
and selected all stars that showed a difference of more than three times
the individual error of the Pulkovo value, but at least 5~km\,s$^{-1}$ to include errors of the Gaia DR2 radial
velocity determination and systematic differences between the two catalogues.
Using this criterion we compiled a final list containing a total of 743 stars showing radial velocity variations of
at least the mentioned threshold.

This list of 743 Hipparcos stars was checked with the SIMBAD database, and we found that
some stars have already been identified as spectroscopic binaries while others are known photometric variable stars.
Variability can also have an effect on variations of the radial velocity of stars.
We eliminated these stars from our catalogue to obtain the final list of 539 stars,
which is presented in the large Table~\ref{tab:pulkovo} that can be found
in \ref{sec:tab} and is also available online as a CSV file on
the website \url{http://www.astro.ugto.mx/~dennis/binary_candidates/}.
The first column gives the Hipparcos (HIP) number of the star and the second column the
difference in the radial velocity (in km\,s$^{-1}$) between the Pulkovo and the Gaia DR2 catalogue measurements.
In this list we found 72 stars with a radial velocity difference of more than 50~km\,s$^{-1}$,
and 42 stars show a variation of more than 100~km\,s$^{-1}$ between the two catalogues.
The largest variation was found for the star HIP~85690 that shows a
difference in the radial velocity of 253.17~km\,s$^{-1}$.

\begin{figure}[t]
    \centerline{\includegraphics[width=0.5\textwidth]{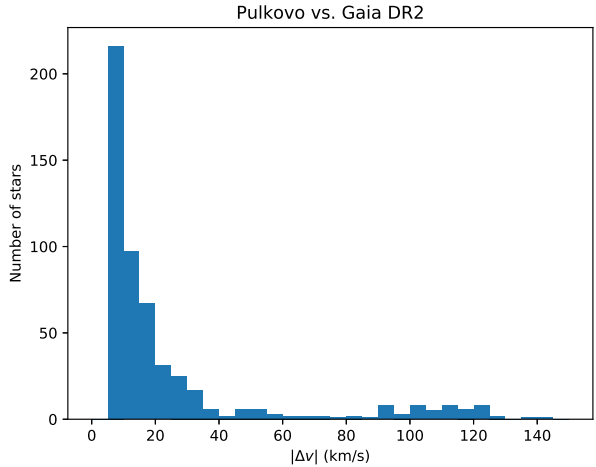}}
        \caption{Distribution of the stars with radial velocity differences ($|\Delta v|$) between the Pulkovo and the Gaia DR2 catalogues
         of $>5$~km\,s$^{-1}$ and at least three times the Pulkovo error.
        \label{figp}}
\end{figure}

A histogram of the radial velocity differences between the Pulkovo and the Gaia DR2 catalogues
is presented in Figure~\ref{figp}. 
There exists a small overdensity in the range of 100 to 130~km\,s$^{-1}$. We
revised all stars with a radial velocity variation of more than 100~km\,s$^{-1}$ and found
that 37 of these 42 stars are classified as either O or B spectral type stars. The radial velocity 
of these types of stars is usually difficult to determine, so that this bump surely
comes from imprecise values of the radial velocities in both catalogues.
The detection rate of spectroscopic binaries in the Pulkovo catalogue is 3.25\%, which is
slightly higher than for the K-M dwarf catalogue.

%
\subsection{Radial velocity experiment (RAVE) catalogue}

The last catalogue for our comparison with the Gaia DR2 radial velocity measurements
was the fifth data release of the Radial Velocity Experiment (RAVE) project \citep{rave}. 
This catalogue is the largest available catalogue of measured stellar radial velocities.
The objects in the catalogue are part of a survey of stars with a magnitude limit of $9<I<12$~mag.
The radial velocities are determined using medium resolution spectra ($R\approx 7500$)
in the Ca~II triplet region in the wavelength range from $8410$ to $8795$~\AA.
The fifth data release of the RAVE catalogue contains 520,701 radial velocity measurements for 457,588 stars.

We used the cross match list of RAVE DR5 sources in the Gaia DR2 database.
As a result, we obtained a list of 456,2913 observations for 401,719 stars that have radial velocity measurements in both catalogues,
meaning that many of the stars in the RAVE catalogue also have Gaia DR2 measurements of the radial velocity. 
The next step was to search for stars that show differences in the radial velocity between the two catalogues.
We used a threshold for the difference of at least 5~km\,s$^{-1}$ and three times the individual error of the RAVE measurement
in radial velocity.
We found a total of 37,073 measurements for 34,896 stars that show variations in the radial velocity between the Gaia DR2 and RAVE
catalogue. 
Eliminating already detected spectroscopic binary stars and photometric variable stars we
obtained a final list of 34,691 stars with 36,844 individual radial velocity measurements.
This corresponds to roughly 8.64\% of the stars of the RAVE catalogue.
We discovered that 4351 stars show a variation of more than 50~km\,s$^{-1}$ in the radial velocity.
A total of 3415 stars with a difference of more that 100~km\,s$^{-1}$ in the radial velocity between
the two catalogues were found. 

However, revising the RAVE catalogue of spectroscopic binary candidate stars we identified
some doubtful cases. The RAVE object with the ID~J063931.6-300221, which is the star with the largest difference in the radial velocity,
shows in the Gaia DR2 database a radial velocity of 31.28~km\,s$^{-1}$ while the RAVE catalogue measurement indicates $-2938.4$~km\,s$^{-1}$.
This difference is extremely high and unlikely to be correct. 
In fact, there are several stars in the RAVE catalogue with very high radial velocities.
We found 1835 measurements that show a radial velocity larger than 400~km\,s$^{-1}$, which
is larger than the escape velocity of our Galaxy. 
It would be good to obtain further observations of the radial velocity of these respective
stars to check the RAVE measurements.

\begin{figure}[t]
    \centerline{\includegraphics[width=0.5\textwidth]{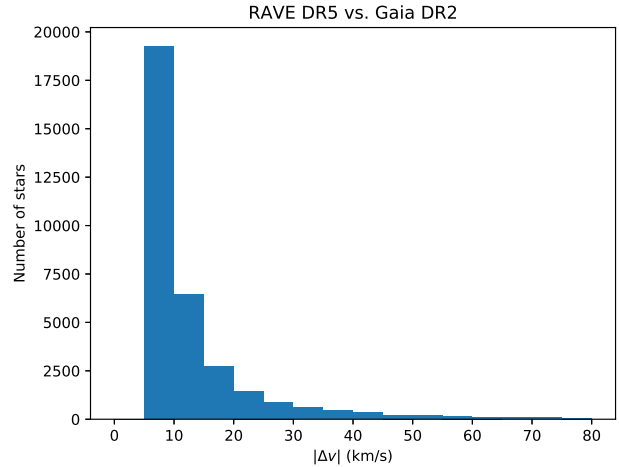}}
        \caption{Distribution of the radial velocity differences ($|\Delta v|$) for the comparison between the RAVE and the Gaia DR2 catalogues.
        \label{fig1}}
\end{figure}

\begin{figure}[t]
    \centerline{\includegraphics[width=0.5\textwidth]{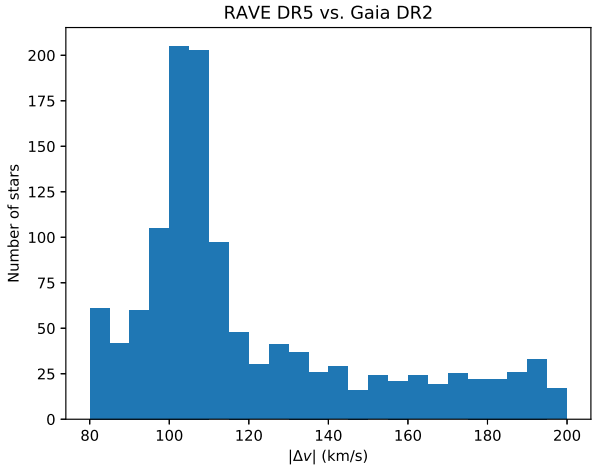}}
        \caption{Distribution of the radial velocity difference ($|\Delta v|$) between the RAVE and the Gaia DR2 catalogues in the range
        from 80 to~200~km\,s$^{-1}$ showing a clear overdensity between 100 and 110~km\,s$^{-1}$.\label{fig2}}
\end{figure}

Since the complete list of the RAVE stars that were added to the spectroscopic binary candidate star catalogue 
contains 36,844 observations, it is too large to present it in this publication. However,
we have made the complete list available online as a CSV file on the website
\url{http://www.astro.ugto.mx/~dennis/binary_candidates/}. Instead, we present some statistics about 
this part of the catalogue. Figure \ref{fig1} shows a histogram of
the number of binary candidate stars found in the RAVE catalogue. Most of the stars have a difference
in radial velocity of between 5 to 10~km\,s$^{-1}$. The higher the radial velocity difference,
the lower the number of stars found. Figure~\ref{fig2} shows the
distribution in the range from 80 to 200~km\,s$^{-1}$. There appears to be a
peak between 100 and 110~km\,s$^{-1}$, whose origin is not clear.
However, there exists a similar
overdensity in the Pulkovo catalogue, where we found that these stars were O or B type stars.
We checked again the spectral type of the stars for the RAVE catalogue, but only very few stars have
a spectral classification so that we could not clearly determine the reason for this peak.

\subsection{Cross matches of stars in both the Pulkovo and RAVE catalogue}

\begin{center}
\begin{table*}[t]%
\centering
\caption{List of stars that show variations of $>5$~km\,s$^{-1}$ and three times the individual error in at least one of the catalogues
of Pulkovo and RAVE, when compared to the Gaia DR2 values.}
\tabcolsep=0pt%
\begin{tabular*}{500pt}{@{\extracolsep\fill}llcccccc@{\extracolsep\fill}}
\toprule
\textbf{HIP} & \textbf{RAVE} & $m_\mathrm{V}$ & \textbf{Spectral type} & \textbf{$v_\mathrm{Pulkovo}$~(km\,s$^{-1}$)} & \textbf{$v_\mathrm{RAVE}$~(km\,s$^{-1}$)}  & \textbf{Epoch$_\mathrm{RAVE}$~(MJD)} &  \textbf{$v_\mathrm{Gaia}$~(km\,s$^{-1}$)} \\
\midrule
3060   &  J003850.3-185205   &  8.29   & F7V   & $ 23.9 \pm 0.5 $ & $22.22 \pm 0.64$ &54348.652 & $16.59 \pm 2.46$ \\
3213   &  J004052.6-512249   & 11.225  & F7Vwl & $ 89.3  \pm 3.4 $ & $123.91 \pm 1.73$ &55044.765 &  $123.25 \pm 0.46$    \\
13749  &  J025700.3-302438   & 10.312  & G0    & $ 104.7 \pm 0.3 $ & $ 99.69 \pm 0.74 $ & 54343.739 & $ 104.91 \pm 0.46 $    \\
24742  &  J051819.7-485213   & 10.619  & F5    & $ 307.9 \pm 2.8 $ & $ 301.82 \pm0.76  $ &53310.716 &  $ 315.75 \pm 5.53 $    \\ 
31809  &  J063901.7-350811   &  7.99   & G8IV  & $ 33.3 \pm 4.2 $ & $ 39.74 \pm 0.99 $ &54462.627 &  $ 31.79 \pm 0.81 $    \\
35424  &  J071849.1-563840   & 9.592   & G8IV/V& $ 62.7 \pm 0.4 $ & $ 52.44 \pm 0.93 $ & 54834.607 &  $ 62.42 \pm 0.37 $    \\ 
35965  &  J072448.3-511939   & 10.18   & K0    & $ 68.0 \pm 4.7 $ & $ 56.50 \pm 1.41 $ &55274.379 &  $ 63.85 \pm 0.24 $    \\
37981  &  J074705.5-275432   &  9.30   &B1/2II/III& $43.0 \pm 4.6$  & $49.59 \pm 49.71$   & 53391.617 & $157.77 \pm 7.0$ \\
       &                     &         &         &        &  $49.55\pm 9.11$ & 53398.625 &   \\
42496  &  J083951.0-660417   & 9.70    & K0/2(V) & $ 23.6 \pm 0.4 $  & $ 25.92 \pm 0.87 $ & 54936.393 &  $  20.76 \pm 0.57  $   \\
48962  &  J095917.9-201024   & 8.11    & F5V  & $ 55.3 \pm  0.7 $ & $ 49.51 \pm 0.88 $ & 55211.693  & $ 55.33 \pm 0.76 $    \\
61478  &  J123550.1-313111   & 10.703  & G8:wl & $185.0 \pm 2.8$  & $200.64 \pm 0.45$   & 53156.431 &  $200.22\pm1.31$\\
64256  &  J131013.0-334236	 & 10.860  &        & $48.0 \pm  2.8$  & $33.04\pm 0.53$  & 54559.647 & $31.55\pm 0.58$   \\
66164  &  J133345.4-635259   &  9.21   & G8/K0V & $ 6.9  \pm 0.4 $ & $ 7.93 \pm 1.64 $ &  55357.398 & $ 6.82\pm 0.29$    \\ 
       &                     &         &        &         & $ 12.85 \pm 1.83 $     &  53214.372  &     \\ 
66693  &  J134016.9-601558   &  9.11   & G6/8V  & $ 35.0 \pm 0.3 $ & $ 41.58 \pm 1.90 $   &53214.372 &  $ 35.35\pm 0.19 $    \\
66814  &  J134142.3-621026   &  9.14   & K0V  & $ -4.8 \pm 0.6 $ & $ -3.718 \pm 1.21 $ & 55348.471 &  $ -5.06 \pm 0.43 $    \\ 
       &                     &         &       &           & $ 1.08  \pm 1.50 $  & 53214.372 &   \\ 
67367  &  J134826.8-602751   &  9.86   & K1/2V & $ -13.6 \pm 0.4 $ & $ -8.07 \pm 1.60 $  & 53214.372 & $ -13.67 \pm 0.40 $    \\
       &                     &         &       &            & $-12.75  \pm 0.73 $ & 55348.471 &    \\
68650  &  J140308.6-321401   & 11.038  & K1:wl & $144.0 \pm 2.8$  & $192.87\pm0.45$   & 53183.460 & $192.50\pm0.74$   \\
       &                     &         &        &          &  $194.37 \pm 0.89$  & 54543.691 & \\
71058  &  J143153.5-311202   &  9.678  & GpBa  & $-53.5 \pm 1.6$  & $-59.89\pm 0.85$   &53812.673 & $-56.16\pm2.84$   \\ 
       &                     &         &       &           &  $-70.96 \pm 1.0$  & 54937.571 & \\
       &                     &         &       &           &  $-71.33 \pm 0.80$  & 54938.601 & \\
       &                     &         &       &           &  $-69.87 \pm 0.84$  & 54951.538 & \\
       &                     &         &      &            &  $-69.17 \pm 0.78$  & 55000.494 & \\
       &                     &         &      &            &  $-67.26 \pm 0.70$  & 55043.368 & \\
113915 &  J230406.0-460002   &  8.13   & F6V &$ 44.1 \pm 0.4 $ & $ 51.29 \pm 0.73 $ & 53609.567  & $ 44.41 \pm 0.20 $    \\ 
       &                     &         &      &            & $ 45.38 \pm 0.49 $ & 53609.573 &    \\ 
\bottomrule
\end{tabular*}
\label{tabcross}
\end{table*}
\end{center}

We checked for stars that have been detected as spectroscopic binary candidates in the Pulkovo, RAVE and Gaia catalogues.
Table~\ref{tabcross} presents the 19 stars obtained and their individual measurements reported in all three catalogues.
The $V$-magnitude of the star and its spectral type
taken from SIMBAD are listed in the third and fourth columns. The last columns show the individual measurements
of the radial velocity (in km\,s$^{-1}$) in the Pulkovo ($v_\mathrm{Pulkovo}$),
RAVE ($v_\mathrm{RAVE}$) and Gaia DR2 ($v_\mathrm{Gaia}$) catalogue. We only list
the epochs for the RAVE observations, since the other catalogues do not provide this information.
There are four stars with a $V$-magnitude of $m_\mathrm{V} < 9$~mag that could be easily
observed by a smaller telescope.

For HIP~3060, HIP~37981, and maybe HIP~24742, the Pulkovo and RAVE values are very similar but show a difference when compared to the Gaia DR2 value.
The measurements for HIP~37981 in all three catalogues show very large errors, which may come from the fact that the star is 
of spectral type B1.
There are other stars where the Pulkovo value is different from the similar RAVE and Gaia DR2 values, like for the stars
HIP~3213, HIP~61478, HIP~64256 and HIP~68650. However, for most of the stars the RAVE value or some of the individual measurements
are different to the similar Pulkovo and Gaia DR2 values (HIP~13749, HIP~31809, HIP~35424, HIP~35965, HIP~48962, HIP~66164, HIP~66693, HIP~67367, HIP~113915).

HIP~71058 has many measurements in the RAVE catalogue, and we found two additional measurements in \citet{catchpole77} of 55.0 and 52.1~km\,s$^{-1}$ having an error of 3~km\,s$^{-1}$.
\citet{bond74} published a radial velocity of $61\pm10$~km\,s$^{-1}$.
HIP~71058 is a Barium Star, which are suspected to be part of a binary system \citep{clure80,clure90} making this star an interesting target
for follow-up observations.

%
\section{Conclusions}

In this work, we compared the new second data release of the Gaia mission that contains a database of radial velocity
measurements of about 7.2 million stars with the previously published radial velocity
values of three different catalogues, namely the K-M dwarf catalogue, the Pulkovo compilation and the RAVE catalogue.

We found a total of 35,246 spectroscopic binary candidate stars (SB1) that we compiled into a
catalogue that has been made publicly available in form of CSV files 
at the website \url{http://www.astro.ugto.mx/~dennis/binary_candidates/}.
In the comparison with the K-M dwarf catalogue we identified 16 spectroscopic binary candidate stars
that we could also study in some detail in the literature.
A total of 539 spectroscopic binary candidate stars have been identified in the
Pulkovo compilation of Hipparcos stars. 
Known spectroscopic binaries and variable stars had been eliminated.
The largest contribution to our catalogue comes from the 
RAVE catalogue, where we found 34,896 stars with radial velocity variations of more than three
times the individual RAVE measurement error and with at least a difference of 5~km\,s$^{-1}$.
19 candidate stars have measurements in the three catalogues of Pulkovo, RAVE and Gaia DR2
and should be very interesting for follow-up observations.

The detection rate in the K-M catalogue is just 2.45\%. This is
lower than expected but comes from the fact that the K-M dwarf catalogue
intentionally does not contain known spectroscopic binary stars.
Clear SB1 stars have also been eliminated from the Gaia DR2 catalogue reducing our
detection rate.
The positive detection rate of the Pulkovo catalogue
is 3.5\%, a bit larger. In the RAVE catalogue we obtained a rate
of  8.64\% of the stars as spectroscopic binary candidates.
The distribution of the RV differences between the Pulkovo and RAVE catalogues shows an
overdensity of stars in a range between 100 to 120 km\,s$^{-1}$. In the Pulkovo catalogue
stars in this range of RV differences are all of O and B spectral types.

Our catalogue of spectroscopic binary candidate stars has been compiled for 
searches and study of spectroscopic binaries and multiple stellar systems.
It will be very useful for follow-up observations of radial velocity measurements of the candidate stars
in order to obtain radial velocity curves.
Only a few observations are needed to verify the identification of binarity in the catalogue.

For the future, we are also planning to use our access to the 1.2~m robotic telescope TIGRE \citep{schmitt14}
to obtain high resolution spectra ($R\approx 20,000$) of some of the spectroscopic binary candidate stars
of our catalogue.
Radial velocity measurement with this telescope have already been used to study the
star HD~16673 to obtain its radial velocity curve and to determine the orbital parameters of 
the system \citep{mittag18}.

\section*{Acknowledgments}

This research has been made possible in part by grants of the Universidad de Guanajuato (UG) through
the DAIP project CIIC 021/2018 and also by CONACyT in several bilateral grants. 

This work has made use of data from the European Space Agency (ESA) mission
{\it Gaia} (\url{https://www.cosmos.esa.int/gaia}), processed by the {\it Gaia}
Data Processing and Analysis Consortium (DPAC,
\url{https://www.cosmos.esa.int/web/gaia/dpac/consortium}). Funding for the DPAC
has been provided by national institutions, in particular the institutions
participating in the {\it Gaia} Multilateral Agreement.

This research has made use of the Pulkovo Compilation of Radial Velocities for 35495 Hipparcos Stars and
the catalogue of Radial velocities of K-M dwarfs that are publicly available
using the VizieR catalogue \citep{vizier} access tool, CDS, Strasbourg, France. 

Funding for RAVE has been provided by: the Australian Astronomical Observatory; 
the Leibniz-Institut fuer Astrophysik Potsdam (AIP); the Australian National University;
the Australian Research Council; the French National Research Agency; the German Research Foundation (SPP 1177 and SFB 881);
the European Research Council (ERC-StG 240271 Galactica); the Istituto Nazionale di Astrofisica at Padova; The Johns Hopkins University;
the National Science Foundation of the USA (AST-0908326); the W. M. Keck foundation; the Macquarie University;
the Netherlands Research School for Astronomy; the Natural Sciences and Engineering Research Council of Canada; 
the Slovenian Research Agency; the Swiss National Science Foundation; the Science \& Technology Facilities Council of the UK;
Opticon; Strasbourg Observatory; and the Universities of Groningen, Heidelberg and Sydney. The RAVE web site is at \url{https://www.rave-survey.org}.

This research has made use of the SIMBAD database, operated at CDS, Strasbourg, France.

\bibliography{all}

\begin{thebibliography}{}

\bibitem [\protect \citeauthoryear {%
{Bond}%
}{%
{Bond}%
}{%
{\protect \APACyear {1974}}%
}]{%
bond74}
\APACinsertmetastar {%
bond74}%
\begin{APACrefauthors}%
{Bond}, H\BPBI E.%
\end{APACrefauthors}%
\unskip\
\newblock
\APACrefYearMonthDay{1974}{{\APACmonth{11}}}{},
\newblock
\unskip
\newblock
\APACjournalVolNumPages{\apj}{194}{}{95-107}.
\newblock
\begin{APACrefDOI} \doi{10.1086/153227} \end{APACrefDOI}
\PrintBackRefs{\CurrentBib}

\bibitem [\protect \citeauthoryear {%
{Catchpole}%
, {Robertson}%
\BCBL {}\ \BBA {} {Warren}%
}{%
{Catchpole}%
\ \protect \BOthers {.}}{%
{\protect \APACyear {1977}}%
}]{%
catchpole77}
\APACinsertmetastar {%
catchpole77}%
\begin{APACrefauthors}%
{Catchpole}, R\BPBI M.%
, {Robertson}, B\BPBI S\BPBI C.%
\BCBL {}\ \BBA {} {Warren}, P\BPBI R.%
\end{APACrefauthors}%
\unskip\
\newblock
\APACrefYearMonthDay{1977}{{\APACmonth{11}}}{},
\newblock
\unskip
\newblock
\APACjournalVolNumPages{\mnras}{181}{}{391-404}.
\newblock
\begin{APACrefDOI} \doi{10.1093/mnras/181.3.391} \end{APACrefDOI}
\PrintBackRefs{\CurrentBib}

\bibitem [\protect \citeauthoryear {%
{Chubak}%
\ \BBA {} {Marcy}%
}{%
{Chubak}%
\ \BBA {} {Marcy}%
}{%
{\protect \APACyear {2011}}%
}]{%
chubak11}
\APACinsertmetastar {%
chubak11}%
\begin{APACrefauthors}%
{Chubak}, C.%
\BCBT {}\ \BBA {} {Marcy}, G.%
\end{APACrefauthors}%
\unskip\
\newblock
\APACrefYearMonthDay{2011}{{\APACmonth{01}}}{},
\newblock
{\BBOQ}\APACrefatitle {{Accurate Radial Velocities of 2284 FGKM Stars and 127
  Standards}} {{Accurate Radial Velocities of 2284 FGKM Stars and 127
  Standards}}.{\BBCQ}
\newblock
\BIn{} \APACrefbtitle {American Astronomical Society Meeting Abstracts \#217}
  {American Astronomical Society Meeting Abstracts \#217}\ \BVOL~43,
  \BPG~434.12.
\PrintBackRefs{\CurrentBib}

\bibitem [\protect \citeauthoryear {%
{Cropper}%
\ \protect \BOthers {.}}{%
{Cropper}%
\ \protect \BOthers {.}}{%
{\protect \APACyear {2018}}%
}]{%
gaiarvspectrometer}
\APACinsertmetastar {%
gaiarvspectrometer}%
\begin{APACrefauthors}%
{Cropper}, M.%
, {Katz}, D.%
, {Sartoretti}, P.%
\ et al.\end{APACrefauthors}%
\unskip\
\newblock
\APACrefYearMonthDay{2018}{{\APACmonth{08}}}{},
\newblock
\unskip
\newblock
\APACjournalVolNumPages{\aap}{616}{}{A5}.
\newblock
\begin{APACrefDOI} \doi{10.1051/0004-6361/201832763} \end{APACrefDOI}
\PrintBackRefs{\CurrentBib}

\bibitem [\protect \citeauthoryear {%
{Cui}%
\ \protect \BOthers {.}}{%
{Cui}%
\ \protect \BOthers {.}}{%
{\protect \APACyear {2012}}%
}]{%
cui12}
\APACinsertmetastar {%
cui12}%
\begin{APACrefauthors}%
{Cui}, X\BHBI Q.%
, {Zhao}, Y\BHBI H.%
, {Chu}, Y\BHBI Q.%
\ et al.\end{APACrefauthors}%
\unskip\
\newblock
\APACrefYearMonthDay{2012}{{\APACmonth{09}}}{},
\newblock
\unskip
\newblock
\APACjournalVolNumPages{Research in Astronomy and
  Astrophysics}{12}{}{1197-1242}.
\newblock
\begin{APACrefDOI} \doi{10.1088/1674-4527/12/9/003} \end{APACrefDOI}
\PrintBackRefs{\CurrentBib}

\bibitem [\protect \citeauthoryear {%
{Dawson}%
\ \BBA {} {De Robertis}%
}{%
{Dawson}%
\ \BBA {} {De Robertis}%
}{%
{\protect \APACyear {2005}}%
}]{%
dawson05}
\APACinsertmetastar {%
dawson05}%
\begin{APACrefauthors}%
{Dawson}, P\BPBI C.%
\BCBT {}\ \BBA {} {De Robertis}, M\BPBI M.%
\end{APACrefauthors}%
\unskip\
\newblock
\APACrefYearMonthDay{2005}{{\APACmonth{01}}}{},
\newblock
\unskip
\newblock
\APACjournalVolNumPages{Publications of the Astronomical Society of the
  Pacific}{117}{}{1-12}.
\newblock
\begin{APACrefDOI} \doi{10.1086/427790} \end{APACrefDOI}
\PrintBackRefs{\CurrentBib}

\bibitem [\protect \citeauthoryear {%
{Gaia Collaboration}%
\ \protect \BOthers {.}}{%
{Gaia Collaboration}%
\ \protect \BOthers {.}}{%
{\protect \APACyear {2018}}%
}]{%
gaiadr2}
\APACinsertmetastar {%
gaiadr2}%
\begin{APACrefauthors}%
{Gaia Collaboration}%
, {Brown}, A\BPBI G\BPBI A.%
, {Vallenari}, A.%
\ et al.\end{APACrefauthors}%
\unskip\
\newblock
\APACrefYearMonthDay{2018}{{\APACmonth{08}}}{},
\newblock
\unskip
\newblock
\APACjournalVolNumPages{\aap}{616}{}{A1}.
\newblock
\begin{APACrefDOI} \doi{10.1051/0004-6361/201833051} \end{APACrefDOI}
\PrintBackRefs{\CurrentBib}

\bibitem [\protect \citeauthoryear {%
{Gaia Collaboration}%
\ \protect \BOthers {.}}{%
{Gaia Collaboration}%
\ \protect \BOthers {.}}{%
{\protect \APACyear {2016}}%
}]{%
gaiamission}
\APACinsertmetastar {%
gaiamission}%
\begin{APACrefauthors}%
{Gaia Collaboration}%
, {Prusti}, T.%
, {de Bruijne}, J\BPBI H\BPBI J.%
\ et al.\end{APACrefauthors}%
\unskip\
\newblock
\APACrefYearMonthDay{2016}{{\APACmonth{11}}}{},
\newblock
\unskip
\newblock
\APACjournalVolNumPages{\aap}{595}{}{A1}.
\newblock
\begin{APACrefDOI} \doi{10.1051/0004-6361/201629272} \end{APACrefDOI}
\PrintBackRefs{\CurrentBib}

\bibitem [\protect \citeauthoryear {%
{Gilmore}%
\ \protect \BOthers {.}}{%
{Gilmore}%
\ \protect \BOthers {.}}{%
{\protect \APACyear {2012}}%
}]{%
gilmore12}
\APACinsertmetastar {%
gilmore12}%
\begin{APACrefauthors}%
{Gilmore}, G.%
, {Randich}, S.%
, {Asplund}, M.%
\ et al.\end{APACrefauthors}%
\unskip\
\newblock
\APACrefYearMonthDay{2012}{{\APACmonth{03}}}{},
\newblock
\unskip
\newblock
\APACjournalVolNumPages{The Messenger}{147}{}{25-31}.
\PrintBackRefs{\CurrentBib}

\bibitem [\protect \citeauthoryear {%
{Gontcharov}%
}{%
{Gontcharov}%
}{%
{\protect \APACyear {2006}}%
}]{%
pulkovo}
\APACinsertmetastar {%
pulkovo}%
\begin{APACrefauthors}%
{Gontcharov}, G\BPBI A.%
\end{APACrefauthors}%
\unskip\
\newblock
\APACrefYearMonthDay{2006}{{\APACmonth{11}}}{},
\newblock
\unskip
\newblock
\APACjournalVolNumPages{Astronomy Letters}{32}{}{759-771}.
\newblock
\begin{APACrefDOI} \doi{10.1134/S1063773706110065} \end{APACrefDOI}
\PrintBackRefs{\CurrentBib}

\bibitem [\protect \citeauthoryear {%
{Hawley}%
, {Gizis}%
\BCBL {}\ \BBA {} {Reid}%
}{%
{Hawley}%
\ \protect \BOthers {.}}{%
{\protect \APACyear {1996}}%
}]{%
hawley96}
\APACinsertmetastar {%
hawley96}%
\begin{APACrefauthors}%
{Hawley}, S\BPBI L.%
, {Gizis}, J\BPBI E.%
\BCBL {}\ \BBA {} {Reid}, I\BPBI N.%
\end{APACrefauthors}%
\unskip\
\newblock
\APACrefYearMonthDay{1996}{{\APACmonth{12}}}{},
\newblock
\unskip
\newblock
\APACjournalVolNumPages{\aj}{112}{}{2799}.
\newblock
\begin{APACrefDOI} \doi{10.1086/118222} \end{APACrefDOI}
\PrintBackRefs{\CurrentBib}

\bibitem [\protect \citeauthoryear {%
{Katz}%
\ \protect \BOthers {.}}{%
{Katz}%
\ \protect \BOthers {.}}{%
{\protect \APACyear {2018}}%
}]{%
gaiarv}
\APACinsertmetastar {%
gaiarv}%
\begin{APACrefauthors}%
{Katz}, D.%
, {Sartoretti}, P.%
, {Cropper}, M.%
\ et al.\end{APACrefauthors}%
\unskip\
\newblock
\APACrefYearMonthDay{2018}{{\APACmonth{04}}}{},
\newblock
\unskip
\newblock
\APACjournalVolNumPages{ArXiv e-prints 1804.09372}{}{}{}.
\PrintBackRefs{\CurrentBib}

\bibitem [\protect \citeauthoryear {%
{Kharchenko}%
, {Scholz}%
, {Piskunov}%
, {R{\"o}ser}%
\BCBL {}\ \BBA {} {Schilbach}%
}{%
{Kharchenko}%
\ \protect \BOthers {.}}{%
{\protect \APACyear {2007}}%
}]{%
kharchenko07}
\APACinsertmetastar {%
kharchenko07}%
\begin{APACrefauthors}%
{Kharchenko}, N\BPBI V.%
, {Scholz}, R\BPBI D.%
, {Piskunov}, A\BPBI E.%
, {R{\"o}ser}, S.%
\BCBL {}\ \BBA {} {Schilbach}, E.%
\end{APACrefauthors}%
\unskip\
\newblock
\APACrefYearMonthDay{2007}{{\APACmonth{11}}}{},
\newblock
\unskip
\newblock
\APACjournalVolNumPages{Astronomische Nachrichten}{328}{}{889}.
\newblock
\begin{APACrefDOI} \doi{10.1002/asna.200710776} \end{APACrefDOI}
\PrintBackRefs{\CurrentBib}

\bibitem [\protect \citeauthoryear {%
{Kirk}%
\ \protect \BOthers {.}}{%
{Kirk}%
\ \protect \BOthers {.}}{%
{\protect \APACyear {2016}}%
}]{%
kepler}
\APACinsertmetastar {%
kepler}%
\begin{APACrefauthors}%
{Kirk}, B.%
, {Conroy}, K.%
, {Pr{\v s}a}, A.%
\ et al.\end{APACrefauthors}%
\unskip\
\newblock
\APACrefYearMonthDay{2016}{{\APACmonth{03}}}{},
\newblock
\unskip
\newblock
\APACjournalVolNumPages{\aj}{151}{}{68}.
\newblock
\begin{APACrefDOI} \doi{10.3847/0004-6256/151/3/68} \end{APACrefDOI}
\PrintBackRefs{\CurrentBib}

\bibitem [\protect \citeauthoryear {%
{Kunder}%
\ \protect \BOthers {.}}{%
{Kunder}%
\ \protect \BOthers {.}}{%
{\protect \APACyear {2017}}%
}]{%
rave}
\APACinsertmetastar {%
rave}%
\begin{APACrefauthors}%
{Kunder}, A.%
, {Kordopatis}, G.%
, {Steinmetz}, M.%
\ et al.\end{APACrefauthors}%
\unskip\
\newblock
\APACrefYearMonthDay{2017}{{\APACmonth{02}}}{},
\newblock
\unskip
\newblock
\APACjournalVolNumPages{\aj}{153}{}{75}.
\newblock
\begin{APACrefDOI} \doi{10.3847/1538-3881/153/2/75} \end{APACrefDOI}
\PrintBackRefs{\CurrentBib}

\bibitem [\protect \citeauthoryear {%
{Majewski}%
, {APOGEE Team}%
\BCBL {}\ \BBA {} {APOGEE-2 Team}%
}{%
{Majewski}%
\ \protect \BOthers {.}}{%
{\protect \APACyear {2016}}%
}]{%
majewski16}
\APACinsertmetastar {%
majewski16}%
\begin{APACrefauthors}%
{Majewski}, S\BPBI R.%
, {APOGEE Team}%
\BCBL {}\ \BBA {} {APOGEE-2 Team}.%
\end{APACrefauthors}%
\unskip\
\newblock
\APACrefYearMonthDay{2016}{{\APACmonth{09}}}{},
\newblock
\unskip
\newblock
\APACjournalVolNumPages{Astronomische Nachrichten}{337}{}{863}.
\newblock
\begin{APACrefDOI} \doi{10.1002/asna.201612387} \end{APACrefDOI}
\PrintBackRefs{\CurrentBib}

\bibitem [\protect \citeauthoryear {%
{Majewski}%
\ \protect \BOthers {.}}{%
{Majewski}%
\ \protect \BOthers {.}}{%
{\protect \APACyear {2017}}%
}]{%
majewski17}
\APACinsertmetastar {%
majewski17}%
\begin{APACrefauthors}%
{Majewski}, S\BPBI R.%
, {Schiavon}, R\BPBI P.%
, {Frinchaboy}, P\BPBI M.%
\ et al.\end{APACrefauthors}%
\unskip\
\newblock
\APACrefYearMonthDay{2017}{{\APACmonth{09}}}{},
\newblock
\unskip
\newblock
\APACjournalVolNumPages{\aj}{154}{}{94}.
\newblock
\begin{APACrefDOI} \doi{10.3847/1538-3881/aa784d} \end{APACrefDOI}
\PrintBackRefs{\CurrentBib}

\bibitem [\protect \citeauthoryear {%
{Maldonado}%
, {Mart{\'\i}nez-Arn{\'a}iz}%
, {Eiroa}%
, {Montes}%
\BCBL {}\ \BBA {} {Montesinos}%
}{%
{Maldonado}%
\ \protect \BOthers {.}}{%
{\protect \APACyear {2010}}%
}]{%
Maldonado10}
\APACinsertmetastar {%
Maldonado10}%
\begin{APACrefauthors}%
{Maldonado}, J.%
, {Mart{\'\i}nez-Arn{\'a}iz}, R\BPBI M.%
, {Eiroa}, C.%
, {Montes}, D.%
\BCBL {}\ \BBA {} {Montesinos}, B.%
\end{APACrefauthors}%
\unskip\
\newblock
\APACrefYearMonthDay{2010}{{\APACmonth{10}}}{},
\newblock
\unskip
\newblock
\APACjournalVolNumPages{\aap}{521}{}{A12}.
\newblock
\begin{APACrefDOI} \doi{10.1051/0004-6361/201014948} \end{APACrefDOI}
\PrintBackRefs{\CurrentBib}

\bibitem [\protect \citeauthoryear {%
{Mason}%
, {Wycoff}%
, {Hartkopf}%
, {Douglass}%
\BCBL {}\ \BBA {} {Worley}%
}{%
{Mason}%
\ \protect \BOthers {.}}{%
{\protect \APACyear {2001}}%
}]{%
mason01}
\APACinsertmetastar {%
mason01}%
\begin{APACrefauthors}%
{Mason}, B\BPBI D.%
, {Wycoff}, G\BPBI L.%
, {Hartkopf}, W\BPBI I.%
, {Douglass}, G\BPBI G.%
\BCBL {}\ \BBA {} {Worley}, C\BPBI E.%
\end{APACrefauthors}%
\unskip\
\newblock
\APACrefYearMonthDay{2001}{{\APACmonth{12}}}{},
\newblock
\unskip
\newblock
\APACjournalVolNumPages{\aj}{122}{}{3466-3471}.
\newblock
\begin{APACrefDOI} \doi{10.1086/323920} \end{APACrefDOI}
\PrintBackRefs{\CurrentBib}

\bibitem [\protect \citeauthoryear {%
{Mason}%
, {Wycoff}%
, {Hartkopf}%
, {Douglass}%
\BCBL {}\ \BBA {} {Worley}%
}{%
{Mason}%
\ \protect \BOthers {.}}{%
{\protect \APACyear {2018}}%
}]{%
wdsc}
\APACinsertmetastar {%
wdsc}%
\begin{APACrefauthors}%
{Mason}, B\BPBI D.%
, {Wycoff}, G\BPBI L.%
, {Hartkopf}, W\BPBI I.%
, {Douglass}, G\BPBI G.%
\BCBL {}\ \BBA {} {Worley}, C\BPBI E.%
\end{APACrefauthors}%
\unskip\
\newblock
\APACrefYearMonthDay{2018}{{\APACmonth{05}}}{},
\newblock
\unskip
\newblock
\APACjournalVolNumPages{VizieR Online Data Catalog}{1}{}{}.
\PrintBackRefs{\CurrentBib}

\bibitem [\protect \citeauthoryear {%
{McClure}%
, {Fletcher}%
\BCBL {}\ \BBA {} {Nemec}%
}{%
{McClure}%
\ \protect \BOthers {.}}{%
{\protect \APACyear {1980}}%
}]{%
clure80}
\APACinsertmetastar {%
clure80}%
\begin{APACrefauthors}%
{McClure}, R\BPBI D.%
, {Fletcher}, J\BPBI M.%
\BCBL {}\ \BBA {} {Nemec}, J\BPBI M.%
\end{APACrefauthors}%
\unskip\
\newblock
\APACrefYearMonthDay{1980}{{\APACmonth{05}}}{},
\newblock
\unskip
\newblock
\APACjournalVolNumPages{\apjl}{238}{}{L35-L38}.
\newblock
\begin{APACrefDOI} \doi{10.1086/183252} \end{APACrefDOI}
\PrintBackRefs{\CurrentBib}

\bibitem [\protect \citeauthoryear {%
{McClure}%
\ \BBA {} {Woodsworth}%
}{%
{McClure}%
\ \BBA {} {Woodsworth}%
}{%
{\protect \APACyear {1990}}%
}]{%
clure90}
\APACinsertmetastar {%
clure90}%
\begin{APACrefauthors}%
{McClure}, R\BPBI D.%
\BCBT {}\ \BBA {} {Woodsworth}, A\BPBI W.%
\end{APACrefauthors}%
\unskip\
\newblock
\APACrefYearMonthDay{1990}{{\APACmonth{04}}}{},
\newblock
\unskip
\newblock
\APACjournalVolNumPages{\apj}{352}{}{709-723}.
\newblock
\begin{APACrefDOI} \doi{10.1086/168573} \end{APACrefDOI}
\PrintBackRefs{\CurrentBib}

\bibitem [\protect \citeauthoryear {%
{Mittag}%
, {Hempelmann}%
, {Fuhrmeister}%
, {Czesla}%
\BCBL {}\ \BBA {} {Schmitt}%
}{%
{Mittag}%
\ \protect \BOthers {.}}{%
{\protect \APACyear {2018}}%
}]{%
mittag18}
\APACinsertmetastar {%
mittag18}%
\begin{APACrefauthors}%
{Mittag}, M.%
, {Hempelmann}, A.%
, {Fuhrmeister}, B.%
, {Czesla}, S.%
\BCBL {}\ \BBA {} {Schmitt}, J\BPBI H\BPBI M\BPBI M.%
\end{APACrefauthors}%
\unskip\
\newblock
\APACrefYearMonthDay{2018}{{\APACmonth{01}}}{},
\newblock
\unskip
\newblock
\APACjournalVolNumPages{AN}{339}{}{53}.
\newblock
\begin{APACrefDOI} \doi{10.1002/asna.201713367} \end{APACrefDOI}
\PrintBackRefs{\CurrentBib}

\bibitem [\protect \citeauthoryear {%
{Ochsenbein}%
, {Bauer}%
\BCBL {}\ \BBA {} {Marcout}%
}{%
{Ochsenbein}%
\ \protect \BOthers {.}}{%
{\protect \APACyear {2000}}%
}]{%
vizier}
\APACinsertmetastar {%
vizier}%
\begin{APACrefauthors}%
{Ochsenbein}, F.%
, {Bauer}, P.%
\BCBL {}\ \BBA {} {Marcout}, J.%
\end{APACrefauthors}%
\unskip\
\newblock
\APACrefYearMonthDay{2000}{{\APACmonth{04}}}{},
\newblock
\unskip
\newblock
\APACjournalVolNumPages{Astronomy and Astrophysics Supplement
  Series}{143}{}{23-32}.
\newblock
\begin{APACrefDOI} \doi{10.1051/aas:2000169} \end{APACrefDOI}
\PrintBackRefs{\CurrentBib}

\bibitem [\protect \citeauthoryear {%
{Perryman}%
\ \protect \BOthers {.}}{%
{Perryman}%
\ \protect \BOthers {.}}{%
{\protect \APACyear {1997}}%
}]{%
hipparcos1}
\APACinsertmetastar {%
hipparcos1}%
\begin{APACrefauthors}%
{Perryman}, M\BPBI A\BPBI C.%
, {Lindegren}, L.%
, {Kovalevsky}, J.%
\ et al.\end{APACrefauthors}%
\unskip\
\newblock
\APACrefYearMonthDay{1997}{{\APACmonth{07}}}{},
\newblock
\unskip
\newblock
\APACjournalVolNumPages{\aap}{323}{}{L49-L52}.
\PrintBackRefs{\CurrentBib}

\bibitem [\protect \citeauthoryear {%
{Pourbaix}%
\ \protect \BOthers {.}}{%
{Pourbaix}%
\ \protect \BOthers {.}}{%
{\protect \APACyear {2004}}%
}]{%
sb9}
\APACinsertmetastar {%
sb9}%
\begin{APACrefauthors}%
{Pourbaix}, D.%
, {Tokovinin}, A\BPBI A.%
, {Batten}, A\BPBI H.%
\ et al.\end{APACrefauthors}%
\unskip\
\newblock
\APACrefYearMonthDay{2004}{{\APACmonth{09}}}{},
\newblock
\unskip
\newblock
\APACjournalVolNumPages{\aap}{424}{}{727-732}.
\newblock
\begin{APACrefDOI} \doi{10.1051/0004-6361:20041213} \end{APACrefDOI}
\PrintBackRefs{\CurrentBib}

\bibitem [\protect \citeauthoryear {%
{Sartoretti}%
\ \protect \BOthers {.}}{%
{Sartoretti}%
\ \protect \BOthers {.}}{%
{\protect \APACyear {2018}}%
}]{%
sartoretti18}
\APACinsertmetastar {%
sartoretti18}%
\begin{APACrefauthors}%
{Sartoretti}, P.%
, {Katz}, D.%
, {Cropper}, M.%
\ et al.\end{APACrefauthors}%
\unskip\
\newblock
\APACrefYearMonthDay{2018}{{\APACmonth{08}}}{},
\newblock
\unskip
\newblock
\APACjournalVolNumPages{\aap}{616}{}{A6}.
\newblock
\begin{APACrefDOI} \doi{10.1051/0004-6361/201832836} \end{APACrefDOI}
\PrintBackRefs{\CurrentBib}

\bibitem [\protect \citeauthoryear {%
{Schmitt}%
\ \protect \BOthers {.}}{%
{Schmitt}%
\ \protect \BOthers {.}}{%
{\protect \APACyear {2014}}%
}]{%
schmitt14}
\APACinsertmetastar {%
schmitt14}%
\begin{APACrefauthors}%
{Schmitt}, J\BPBI H\BPBI M\BPBI M.%
, {Schr{\"o}der}, K\BHBI P.%
, {Rauw}, G.%
\ et al.\end{APACrefauthors}%
\unskip\
\newblock
\APACrefYearMonthDay{2014}{{\APACmonth{10}}}{},
\newblock
\unskip
\newblock
\APACjournalVolNumPages{AN}{335}{}{787}.
\newblock
\begin{APACrefDOI} \doi{10.1002/asna.201412116} \end{APACrefDOI}
\PrintBackRefs{\CurrentBib}

\bibitem [\protect \citeauthoryear {%
{Sperauskas}%
\ \protect \BOthers {.}}{%
{Sperauskas}%
\ \protect \BOthers {.}}{%
{\protect \APACyear {2016}}%
}]{%
kmdwarfs}
\APACinsertmetastar {%
kmdwarfs}%
\begin{APACrefauthors}%
{Sperauskas}, J.%
, {Barta{\v s}i{\= u}t{\.e}}, S.%
, {Boyle}, R\BPBI P.%
, {Deveikis}, V.%
, {Raudeli{\= u}nas}, S.%
\BCBL {}\ \BBA {} {Upgren}, A\BPBI R.%
\end{APACrefauthors}%
\unskip\
\newblock
\APACrefYearMonthDay{2016}{{\APACmonth{12}}}{},
\newblock
\unskip
\newblock
\APACjournalVolNumPages{\aap}{596}{}{A116}.
\newblock
\begin{APACrefDOI} \doi{10.1051/0004-6361/201527850} \end{APACrefDOI}
\PrintBackRefs{\CurrentBib}

\bibitem [\protect \citeauthoryear {%
{Taylor}%
, {Harvin}%
\BCBL {}\ \BBA {} {McAlister}%
}{%
{Taylor}%
\ \protect \BOthers {.}}{%
{\protect \APACyear {2003}}%
}]{%
chara}
\APACinsertmetastar {%
chara}%
\begin{APACrefauthors}%
{Taylor}, S\BPBI F.%
, {Harvin}, J\BPBI A.%
\BCBL {}\ \BBA {} {McAlister}, H\BPBI A.%
\end{APACrefauthors}%
\unskip\
\newblock
\APACrefYearMonthDay{2003}{{\APACmonth{05}}}{},
\newblock
\unskip
\newblock
\APACjournalVolNumPages{\pasp}{115}{}{609-617}.
\newblock
\begin{APACrefDOI} \doi{10.1086/374759} \end{APACrefDOI}
\PrintBackRefs{\CurrentBib}

\bibitem [\protect \citeauthoryear {%
{van Leeuwen}%
}{%
{van Leeuwen}%
}{%
{\protect \APACyear {2007}}%
}]{%
hipparcos2}
\APACinsertmetastar {%
hipparcos2}%
\begin{APACrefauthors}%
{van Leeuwen}, F.%
\end{APACrefauthors}%
\unskip\
\newblock
\APACrefYearMonthDay{2007}{{\APACmonth{11}}}{},
\newblock
\unskip
\newblock
\APACjournalVolNumPages{\aap}{474}{}{653-664}.
\newblock
\begin{APACrefDOI} \doi{10.1051/0004-6361:20078357} \end{APACrefDOI}
\PrintBackRefs{\CurrentBib}

\bibitem [\protect \citeauthoryear {%
{Wenger}%
\ \protect \BOthers {.}}{%
{Wenger}%
\ \protect \BOthers {.}}{%
{\protect \APACyear {2000}}%
}]{%
simbad}
\APACinsertmetastar {%
simbad}%
\begin{APACrefauthors}%
{Wenger}, M.%
, {Ochsenbein}, F.%
, {Egret}, D.%
\ et al.\end{APACrefauthors}%
\unskip\
\newblock
\APACrefYearMonthDay{2000}{{\APACmonth{04}}}{},
\newblock
\unskip
\newblock
\APACjournalVolNumPages{\aaps}{143}{}{9-22}.
\newblock
\begin{APACrefDOI} \doi{10.1051/aas:2000332} \end{APACrefDOI}
\PrintBackRefs{\CurrentBib}

\end{thebibliography}

\appendix

\section{Spectroscopic binary candidate stars in the K-M dwarf catalogue}
\label{sec:km}

\noindent\textbf{GJ 9038B} There is a difference of 4.05~km\,s$^{-1}$ in the radial velocity when compared to the Gaia measurement.
However, the individual measurements in the K-M catalogue vary from -2.2 to 0.4~km\,s$^{-1}$ with an approximate error of 1~km\,s$^{-1}$.
Thus, this detection is still not very clear, further observations are required, and
the presence of the near-by star GJ~9038A at a separation of about 6.15~arcseconds should also be mentioned.
GJ~9038B has a parallax of $37.4109 \pm 0.1135$~mas which corresponds to a distance of about 26.7~pc. \\

\noindent\textbf{HIP 14165} There exists a difference of 12~km\,s$^{-1}$ between the radial velocity measurement
of Gaia DR2 and of the K-M dwarf catalogue, which only has one individual measurement.
Another measurement of the radial velocity with a value of $41\pm 10 $~km\,s$^{-1}$ can be found in \citet{hawley96},
which has a very high error, though.
The error of the Gaia DR2 measurement is also quite large, but that could indicate a variation of the radial velocity
showing that this star is likely a spectroscopic binary.\\

\noindent\textbf{HIP 14593} For this star a difference in the radial velocity measurements of about 12~km\,s$^{-1}$ between
the two catalogues has been obtained.
In the K-M dwarf catalogue, only one individual measurement has been obtained. This star is very likely to be a spectroscopic binary,
but further observations are required to confirm this.\\

\noindent\textbf{HIP 34361} This star shows a difference in the radial velocity of almost 44~km\,s$^{-1}$ between
the two catalogues. In the K-M dwarf catalogue only one individual measurement has been reported. This star is also a very
good spectroscopic binary star candidate, thus further observations are encouraged.\\

\noindent\textbf{HIP 40724} This is a very interesting star. Although there exists a radial velocity 
difference of only 6.2~km\,s$^{-1}$, other measurements of \citet{kharchenko07} report a radial velocity of $55.0\pm 10$~km\,s$^{-1}$.
However, the two individual measurements in the K-M dwarfs catalogue
of 89.2 and 87.1~km\,s$^{-1}$ with an error of 0.5~km\,s$^{-1}$ are a bit inconclusive. The error of the Gaia DR2 value is
large, indicating a possible binary star. \\

\noindent\textbf{TYC 1379-91-1} This star shows a difference in the radial velocity of almost 54~km\,s$^{-1}$.
However, the four individual observations of the K-M dwarf catalogue do not show a variation in the radial velocity.\\

\noindent\textbf{HIP 44109} There exists just one individual observation in the K-M dwarf catalogue and the difference with the Gaia
observation is 4.3~km\,s$^{-1}$. The error of the Gaia DR2 value is quite high, which could indicate a possible
variation in the radial velocity. There exists a measurement published in the catalogue of \citet{kharchenko07} that
reports a value of $36 \pm 10$~km\,s$^{-1}$, which is slightly different, but still has a quite large error.\\

\noindent\textbf{HIP 46662} There is a difference of 6.1~km\,s$^{-1}$ in the radial velocity
measurements between the two catalogues. The
individual observations in the K-M dwarf catalogue show a change from $-50.8$ to $-51.7$~km\,s$^{-1}$. The error
of the Gaia measurement is larger than 1~km\,s$^{-1}$.\\

\noindent\textbf{HIP 49969} The difference in the radial velocity is 5.7~km\,s$^{-1}$. The two individual observations
in the K-M catalogue are almost the same (26.1 and 25.7~km\,s$^{-1}$). There exists an observation in the Pulkovo catalogue \citep{pulkovo}
with a value of $26.8 \pm 0.8$~km\,s$^{-1}$, which is close to the K-M catalogue value. The value of the Gaia DR2 is larger and
also has a larger error of more than 1~km\,s$^{-1}$.\\

\noindent\textbf{HIP 58099} The difference between the two catalogues is 4.1~km\,s$^{-1}$. The two individual
observations of the K-M dwarf catalogue are very close. \citet{kharchenko07} reports a 
radial velocity of $-7.0\pm 10$~km\,s$^{-1}$ which shows a difference but with a very large error.
There remain doubts that this is really a spectroscopic binary, so further observations with measurements
of the radial velocity would be very helpful.\\

\noindent\textbf{HIP 59000} There exists a difference of 21~km\,s$^{-1}$ in the radial velocity
between the Gaia DR2 and K-M dwarf catalogue measurements. This stars has three measurements of the radial velocity reported 
by \citet{Maldonado10} of 12.2, 33.5 and 52.8~km\,s$^{-1}$. This star is clearly a spectroscopic binary.\\

\noindent\textbf{TYC 4558-1019-1} This star shows a difference of 83~km\,s$^{-1}$ in the radial velocity
measurements, but the individual observations in the K-M dwarf catalogue do not show variations. 
However, the error of the Gaia measurement is larger than 1~km\,s$^{-1}$.\\

\noindent\textbf{GJ 4327} This star has a difference of about 20~km\,s$^{-1}$ in the radial velocity between both catalogues.
The two individual measurements of the K-M dwarf catalogue of 14.0 and 11.0~km\,s$^{-1}$ show a small difference.
Another observation of this star by \citet{dawson05} measured a different value of 0.58~km\,s$^{-1}$ for the radial
velocity.
This is very likely a spectroscopic binary star, also considering the large error of the Gaia DR2 value.\\

\noindent\textbf{HIP 115004} There is a difference in the radial velocity of 7.5~km\,s$^{-1}$.
Individual observations in the K-M dwarf catalogue range from 25.5 to 27.1~km\,s$^{-1}$. There exists a further
reported observation with value of 26.9~km\,s$^{-1}$ by \citet{chubak11}. \\

\noindent\textbf{HIP 116936} The difference in radial velocity was found to be about 11~km\,s$^{-1}$ between
the two catalogues. 
The three individual measurements in the K-M dwarf catalogue do not show a variation in the radial velocity. 
However, the Gaia DR2 value has a large error of about 2.4~km\,s$^{-1}$.
A further measurement reported in the Pulkovo catalogue \citep{pulkovo} gives a value of $25.0 \pm 2.6$~km\,s$^{-1}$.
This value is different to the K-M dwarf catalogue and close to the Gaia DR2 measurement. \\

\section{Pulkovo Table}
\label{sec:tab}

Table \ref{tab:pulkovo} presents the final list of 539 spectroscopic binary candidate stars
that show a variation of at least 5~km\,s$^{-1}$ and more than three times the error of the measurement of the
Pulkovo catalogue value in the radial velocity and that are not known as either spectroscopic binaries or variable stars.
The first column gives the HIP~number of the star while the second states the radial velocity difference ($v_\mathrm{Pulkovo}-v_\mathrm{Gaia}$) in km\,s$^{-1}$.

\begin{center}
\begin{table*}[t]%
\centering
\caption{List of stars of the Pulkovo catalogue that show a variation in the radial velocity
$\Delta v = v_\mathrm{Pulkovo} - v_\mathrm{Gaia}$ (in km\,s$^{-1}$)
of more than 5~km\,s$^{-1}$ and three times the error when compared with the Gaia DR2 catalogue.}
\tabcolsep=0pt%
\begin{tabular*}{500pt}{@{\extracolsep\fill}lc|lc|lc|lc|lc|lc@{\extracolsep\fill}}
\toprule
\textbf{HIP} & \textbf{$\Delta v$}  & \textbf{HIP} & \textbf{$\Delta v$}  & \textbf{HIP} & \textbf{$\Delta v$}   & \textbf{HIP} & 
\textbf{$\Delta v$} & \textbf{HIP} & \textbf{$\Delta v$} & \textbf{HIP} & \textbf{$\Delta v$} \\
\midrule
23 & 10.86 & 8018 & -6.98 & 18600 & 9.15 & 28498 & -13.58 & 38179 & 9.91 & 51150 & -119.98 \\
56 & 58.7 & 8239 & -114.56 & 18745 & -10.0 & 28663 & 30.23 & 38567 & -19.77 & 51240 & -6.41 \\
88 & -6.7 & 8592 & 27.23 & 19014 & -117.34 & 28846 & -8.58 & 38693 & -15.66 & 52584 & 11.6 \\
319 & -103.24 & 8807 & -15.2 & 19023 & 27.26 & 28904 & -113.24 & 38915 & -56.03 & 52845 & -12.51 \\
324 & 7.22 & 9013 & 7.9 & 19027 & 7.29 & 28933 & -6.03 & 39337 & -25.11 & 53155 & 6.24 \\
343 & -52.85 & 9056 & -9.41 & 19050 & -15.62 & 29166 & 18.5 & 39819 & -5.88 & 53522 & -21.49 \\
431 & -7.14 & 9059 & -13.77 & 19089 & -5.62 & 29242 & -7.67 & 40479 & 9.64 & 53750 & -5.31 \\
482 & 9.78 & 9249 & 21.41 & 19404 & -114.16 & 29558 & -7.92 & 40597 & 13.25 & 54481 & -22.6 \\
678 & -19.68 & 9272 & -18.73 & 19628 & -24.93 & 29764 & -16.89 & 40868 & -7.07 & 54886 & -6.36 \\
806 & 15.42 & 9409 & 14.69 & 20049 & -31.62 & 29967 & -8.87 & 40913 & -7.86 & 56050 & -121.9 \\
1038 & -5.51 & 9538 & -42.43 & 20150 & 6.63 & 29979 & -103.12 & 41040 & 8.3 & 56318 & -12.18 \\
1250 & 13.63 & 9818 & -14.32 & 20381 & -93.91 & 30003 & 8.31 & 41187 & 17.63 & 56524 & -15.74 \\
1440 & 5.57 & 10060 & -89.24 & 20595 & -13.49 & 30321 & 10.24 & 41443 & -20.22 & 56624 & 9.92 \\
1449 & 27.72 & 10472 & -9.94 & 20984 & 8.57 & 30385 & -16.86 & 41544 & -7.74 & 57160 & -13.46 \\
1594 & -9.47 & 10767 & 5.93 & 20988 & -11.22 & 30403 & -8.68 & 42133 & -5.13 & 57326 & -14.3 \\
1721 & -7.85 & 10973 & 9.22 & 21023 & -7.68 & 30512 & -15.32 & 42587 & -101.07 & 57344 & 6.4 \\
1901 & 17.05 & 11273 & 16.41 & 21388 & -8.48 & 30724 & -8.74 & 42591 & -29.0 & 57353 & -26.58 \\
2045 & -14.13 & 11453 & -10.87 & 21614 & -7.99 & 30779 & 6.5 & 42941 & -135.26 & 57784 & -22.46 \\
2095 & -6.04 & 11734 & 17.4 & 21805 & 6.82 & 31078 & -31.23 & 42950 & 29.75 & 57951 & -39.8 \\
2390 & -84.78 & 11763 & -115.23 & 22145 & -12.73 & 31237 & -5.18 & 43125 & -94.59 & 58208 & -9.54 \\
2597 & 25.22 & 11770 & -7.31 & 22366 & -16.78 & 31542 & -102.26 & 43260 & -109.19 & 58366 & 62.23 \\
2804 & -12.17 & 11797 & 6.58 & 22524 & 11.07 & 31598 & -121.6 & 43406 & 224.53 & 58740 & -13.63 \\
2816 & -94.47 & 12379 & -116.86 & 22672 & -11.24 & 31641 & 15.02 & 43673 & -10.12 & 59002 & 15.61 \\
2905 & -114.52 & 12652 & 7.98 & 22691 & 48.09 & 32754 & 17.14 & 43808 & 7.19 & 59023 & -19.41 \\
3009 & 35.52 & 12653 & -6.41 & 22974 & -20.08 & 32935 & -29.22 & 44107 & 9.21 & 59052 & -22.42 \\
3060 & 7.31 & 12710 & -5.77 & 23886 & -9.91 & 33048 & 6.15 & 44335 & -10.3 & 59078 & -6.1 \\
3213 & -33.95 & 12769 & -106.29 & 24304 & -8.06 & 33494 & 9.95 & 44413 & -72.73 & 59100 & 20.58 \\
3258 & 9.95 & 13326 & 15.79 & 24513 & -14.22 & 33577 & -6.96 & 44553 & 26.27 & 59234 & -15.43 \\
3268 & 38.22 & 13393 & 6.2 & 24548 & -39.39 & 33658 & 78.61 & 44724 & -8.03 & 59270 & -9.96 \\
3547 & 32.84 & 13447 & 15.39 & 24622 & -9.18 & 34089 & -5.69 & 44755 & 5.51 & 59422 & -5.7 \\
3781 & 20.17 & 13891 & -10.47 & 25268 & -11.73 & 34523 & -94.35 & 45033 & -9.76 & 59535 & 9.81 \\
3933 & 42.72 & 14233 & -5.02 & 25305 & 5.61 & 34750 & -7.6 & 45155 & -8.27 & 59557 & 24.49 \\
4531 & 5.71 & 14244 & 15.26 & 25369 & 16.14 & 34795 & 5.79 & 45536 & -18.02 & 59611 & -8.77 \\
4887 & 18.1 & 14630 & 45.33 & 25418 & -13.3 & 35027 & 8.78 & 45595 & -5.87 & 59850 & 8.97 \\
4929 & 22.17 & 15476 & -31.56 & 25498 & -19.77 & 35427 & -5.92 & 45826 & -9.66 & 60365 & -7.82 \\
5367 & 15.33 & 15486 & -6.63 & 25659 & 16.2 & 36233 & 12.68 & 45979 & -7.35 & 60441 & -7.95 \\
5497 & 6.7 & 15779 & 5.14 & 25715 & 22.63 & 36268 & -32.82 & 46011 & 11.34 & 60453 & -8.62 \\
5647 & -14.82 & 15875 & 26.46 & 25947 & -124.0 & 36322 & -23.7 & 46170 & 21.63 & 60468 & 13.09 \\
5660 & -9.09 & 15920 & -22.95 & 25997 & -9.66 & 36480 & -9.67 & 46511 & -23.79 & 60638 & 9.77 \\
5898 & 18.47 & 16174 & -37.77 & 26342 & 8.87 & 36727 & -127.77 & 46802 & 39.98 & 60779 & -7.18 \\
6291 & 17.57 & 16343 & -9.86 & 26550 & -15.55 & 36896 & 6.11 & 47176 & 7.75 & 60993 & -100.61 \\
6490 & -8.4 & 16613 & -8.41 & 26759 & -20.09 & 37127 & -34.25 & 48310 & -13.47 & 61168 & -16.29 \\
6665 & -26.27 & 16714 & 74.01 & 26765 & -13.82 & 37294 & -22.13 & 48329 & -17.27 & 61293 & -53.48 \\
6901 & -8.31 & 16885 & -32.75 & 27153 & -15.17 & 37376 & -12.29 & 49208 & -7.44 & 61314 & -7.52 \\
7067 & 33.92 & 17497 & -11.21 & 27173 & 12.95 & 37383 & -10.41 & 49389 & 8.91 & 61361 & -7.17 \\
7443 & -6.07 & 17726 & -14.2 & 27267 & 26.4 & 37587 & 28.48 & 49874 & -83.31 & 61478 & -15.22 \\
7636 & -8.87 & 17920 & -9.57 & 27498 & 24.4 & 37854 & -9.57 & 50059 & 26.16 & 61666 & 8.99 \\
7841 & 14.89 & 18011 & -10.87 & 27791 & 20.69 & 37945 & -12.36 & 50251 & -6.48 & 61705 & 5.11 \\
7877 & -16.53 & 18050 & -6.3 & 27836 & 28.55 & 37981 & -114.77 & 50272 & -122.06 & 61707 & -9.42 \\
7924 & -11.37 & 18188 & 30.84 & 28356 & -19.6 & 38013 & -16.32 & 50909 & -31.42 & 61772 & -99.82 \\
\bottomrule
\end{tabular*}
\label{tab:pulkovo}
\end{table*}
\end{center}

\setcounter{table}{0}

\begin{center}
\begin{table*}[t]%
\centering
\caption{(continued)}
\tabcolsep=0pt%
\begin{tabular*}{500pt}{@{\extracolsep\fill}lc|lc|lc|lc|lc|lc@{\extracolsep\fill}}
\toprule
\textbf{HIP} & \textbf{$\Delta v$}  & \textbf{HIP} & \textbf{$\Delta v$}  & \textbf{HIP} & \textbf{$\Delta v$}   & \textbf{HIP} & 
\textbf{$\Delta v$} & \textbf{HIP} & \textbf{$\Delta v$} & \textbf{HIP} & \textbf{$\Delta v$} \\
\midrule
61980 & -18.6 & 69064 & 8.99 & 82440 & 25.7 & 93623 & -11.62 & 100988 & 8.82 & 110212 & 20.91 \\
62076 & -6.18 & 70072 & -14.86 & 82740 & 5.27 & 93796 & -123.24 & 101186 & -109.39 & 110221 & 8.12 \\
62178 & -6.98 & 70239 & -5.46 & 82994 & 5.11 & 93929 & -8.77 & 101315 & -34.17 & 110668 & -6.78 \\
62230 & -7.65 & 70344 & -10.3 & 83607 & -6.29 & 94214 & -8.93 & 101747 & 9.22 & 110755 & -18.07 \\
62294 & -9.37 & 70609 & 54.24 & 83767 & -5.99 & 94465 & 15.72 & 101870 & -7.49 & 111273 & -17.45 \\
62298 & -9.89 & 70761 & 8.4 & 84115 & -17.35 & 94598 & -5.79 & 102127 & -5.45 & 111420 & -8.98 \\
62531 & -24.79 & 70857 & -10.26 & 84298 & -18.8 & 94627 & 7.62 & 102289 & -8.52 & 111558 & -6.83 \\
62570 & -28.53 & 70938 & 57.2 & 84588 & -93.67 & 94883 & -13.99 & 102400 & 32.21 & 111568 & 11.61 \\
62591 & -7.37 & 71287 & -13.3 & 85039 & -14.36 & 95326 & 5.76 & 103113 & -19.56 & 111617 & 8.71 \\
62685 & -10.62 & 71602 & -8.09 & 85690 & -253.17 & 95333 & -11.6 & 103130 & 12.87 & 112768 & -12.84 \\
62817 & -8.65 & 71746 & 13.13 & 85815 & -6.32 & 95856 & -5.04 & 103375 & -19.18 & 113444 & 6.42 \\
62832 & -10.0 & 72038 & -9.89 & 86234 & 13.34 & 95996 & 152.78 & 103492 & 8.9 & 113500 & 25.79 \\
62928 & -13.08 & 72382 & -12.28 & 86399 & -5.06 & 96015 & -21.68 & 103639 & 8.1 & 113559 & -26.6 \\
63322 & -34.48 & 72408 & -7.57 & 86732 & -10.47 & 96622 & -5.61 & 104408 & -5.3 & 113999 & -94.29 \\
63433 & 9.11 & 73425 & 30.47 & 86862 & -46.5 & 97189 & -118.34 & 104592 & -6.91 & 114092 & -8.15 \\
63667 & -11.61 & 73614 & -9.05 & 87202 & -7.93 & 97276 & -13.5 & 104681 & 9.46 & 114452 & -7.71 \\
63841 & -8.44 & 73652 & -10.09 & 87247 & -64.4 & 97470 & 23.3 & 104695 & -93.04 & 114725 & -11.0 \\
63910 & -9.87 & 73963 & 8.44 & 87488 & 65.98 & 97665 & -10.62 & 105012 & -12.8 & 114753 & -13.06 \\
63976 & -52.44 & 73991 & 101.45 & 87744 & -14.28 & 97814 & -21.07 & 105514 & 5.91 & 114893 & 21.66 \\
64161 & -17.61 & 74062 & -8.24 & 87999 & 20.98 & 97840 & -21.3 & 105681 & 12.17 & 114990 & -109.34 \\
64256 & 16.45 & 74240 & -15.11 & 88380 & -5.58 & 98116 & -10.62 & 105854 & 5.6 & 115412 & -15.45 \\
64280 & -9.33 & 75132 & 50.36 & 88532 & -117.25 & 98254 & -13.8 & 106042 & -15.13 & 115503 & -6.96 \\
64881 & -6.54 & 75224 & -121.28 & 88851 & -10.28 & 98542 & -7.31 & 106233 & -11.95 & 115836 & -5.75 \\
64938 & -26.67 & 76209 & -26.61 & 89126 & -7.68 & 98658 & 15.94 & 106315 & -9.42 & 115860 & 7.88 \\
65175 & -16.09 & 77110 & -5.73 & 89218 & -109.98 & 98818 & -52.86 & 106440 & -8.22 & 116190 & -12.71 \\
65320 & 15.25 & 77391 & -104.92 & 89294 & -6.42 & 98825 & -10.09 & 106646 & -9.65 & 116471 & 27.65 \\
65458 & -6.89 & 78327 & -5.27 & 89729 & -140.54 & 99312 & -20.88 & 106828 & 10.03 & 116656 & -5.79 \\
66949 & -5.91 & 78881 & -9.43 & 89988 & 13.16 & 99318 & 7.1 & 107134 & 7.35 & 116726 & -18.39 \\
66969 & -6.41 & 79234 & 19.72 & 90199 & 66.2 & 99602 & -7.09 & 107193 & -9.5 & 116728 & 5.56 \\
67071 & -10.35 & 79254 & 11.42 & 90267 & -128.01 & 99607 & -7.94 & 107465 & 13.68 & 116746 & 9.71 \\
67261 & -8.26 & 79382 & -47.29 & 90810 & -8.57 & 99635 & 23.27 & 107688 & -14.25 & 116957 & 9.53 \\
67351 & 6.63 & 80131 & -19.32 & 90936 & -8.65 & 99646 & 14.13 & 108039 & -8.78 & 116973 & 13.38 \\
67676 & -11.01 & 80221 & 16.49 & 90950 & -120.82 & 99725 & 12.45 & 108546 & -111.31 & 117088 & -8.39 \\
67685 & -7.98 & 80323 & -99.23 & 91249 & 26.79 & 99840 & -16.73 & 108868 & -9.83 & 117213 & 7.05 \\
67733 & -14.72 & 80984 & -14.51 & 91822 & -91.91 & 99939 & 5.27 & 109081 & 7.89 & 117749 & -12.25 \\
67790 & -15.74 & 81196 & 15.8 & 91934 & -6.64 & 100118 & -5.23 & 109258 & -7.93 & 118139 & -120.98 \\
67883 & 10.49 & 81378 & 14.24 & 92198 & -98.44 & 100176 & 11.54 & 109307 & 13.28 & 118148 & -11.48 \\
68557 & -6.23 & 81966 & 10.74 & 92223 & -6.21 & 100391 & -114.23 & 109961 & 33.47 & 118230 & -6.58 \\
68650 & -48.5 & 82245 & 5.21 & 92434 & -28.17 & 100404 & 19.85 & 110020 & -113.15 & 118252 & -46.93 \\
68940 & -17.36 & 82281 & -16.65 & 92880 & -11.85 & 100557 & -102.68 & 110053 & 8.66    &    &    \\
\bottomrule
\end{tabular*}
\end{table*}
\end{center}

\end{document}